\documentstyle[prb,aps,eqsecnum]{revtex}  
\begin{document}   
\baselineskip 0.6cm          
\title{ LARGE LOW TEMPERATURE MAGNETORESISTANCE AND MAGNETIC ANOMALIES  IN
Tb$_2$PdSi$_3$ and Dy$_2$PdSi$_3$ }
\author{R. Mallik, E.V. Sampathkumaran and P.L. Paulose}
\address{ Tata Institute of Fundamental Research,  Homi  Bhabha  Road,
Mumbai-400005, INDIA. }
\maketitle

\begin{abstract}
The results of heat-capacity (C), magnetic  susceptibility $(\chi )$,
electrical resistivity $(\rho )$ and magnetoresistance $(\Delta \rho
/\rho)$ measurements on the  compounds, Tb$_2$PdSi$_3$ and Dy$_2$PdSi$_3$,
are reported.  The  results  establish  that  these compounds undergo
long-range magnetic ordering (presumably  with  a  complex magnetic
structure) below (T$_c$=) 23  and 8 K  respectively.  The $\Delta \rho
/\rho $  is negative in the vicinity of T$_c$ and the magnitude grows as
T$_c$  is  approached from higher temperature as in the case of well-known
giant magnetoresistance systems (La manganite based perovskites); this is
attributed to the formation  of  some kind of magnetic polarons. The
magnitude of $\Delta \rho /\rho $ at low temperatures is quite large, for
instance, about 30\%  in  the  presence  of  60  kOe field at 5 K in Dy
sample.
\end{abstract}

\vskip 1cm
Among the ternary rare-earth (R) intermetallic compounds,  only  a  few
series of compounds of the type, R$_2$XY$_3$ (X=  transition  metals;
Y=Si, Ga), crystallizing  in  AlB$_2$-derived  hexagonal structures,  has
been   known [1-8].  Recently, we have been paying special attention [8,
9] to  the series, R$_2$PdSi$_3$ (Ref. 2). We reported the formation of
Ce$_2$PdSi$_3$ (Ref. 8) and Eu$_2$PdSi$_3$ (Ref. 9), which  were otherwise
not  known  earlier.  During  the   course   of investigation of these Pd
compounds,  we  have  noted a  minimum  in the temperature dependent
electrical resistivity $(\rho )$ before long range  magnetic ordering in
the Gd  alloy  (Ref.  9),  mimicking the  behavior  in magnetically
ordering Kondo lattices, as a novel magnetic precursor effect.  We have
proposed  that  possibly  the tendency  for  the localisation  of
conduction  electrons  due  to exchange  interaction (prior  to   magnetic
ordering) is responsible for this behavior. From this result, we inferred
that some kind  of  magnetic polaronic  effects  can  be operative to some
extent,  not   only in semiconductors [10, 11], but also in metallic
systems  [9]  prior  to long range magnetic ordering. In order to look for
such effects in  other compounds containing magnetic-moment carrying R
ions and also as a continuation of our efforts to understand the magnetic
behavior of this class of compounds,  we have undertaken the
investigations on Tb$_2$PdSi$_3$ and Dy$_2$PdSi$_3$.
\par
The samples were prepared by  arc  melting  stoichiometric  amounts  of
constituent elements in an atmosphere  of  argon.  The  molten  ingots
were homogenised at 750 C for  five  days  and  the x-ray  diffraction
patterns confirm  that  these  compounds  crystallize  in  a
AlB$_2$-derived  hexagonal structure [2]. The heat-capacity (C)
measurements  were  performed  in  the temperature interval 3 - 60 K by a
semiadiabatic heat-pulse method [12]. The magnetic susceptibility $(\chi)$
data (2 - 300 K)  were  obtained  employing  a commercial superconducting
quantum interference device in a  magnetic  field (H) of 2 kOe. Ac $\chi$
measurements (ac field 0.8 Oe) were also performed (4.2 - 100 K) at two
different frequencies, 106 and 1090 Hz. The $\rho $ data (2 - 300 K) were
taken by a conventional four-probe method; in addition, the data were
collected in the presence of a  field  of 50 kOe as a function of
temperature (T up to about 100 K) and  also  as  a function of H at
selected temperatures (5, 10 and 35 K);  the  direction of the current is
the same as that of H.
\par
The results of C, $\chi$ and $\rho$ measurements are shown for
Tb$_2$PdSi$_3$  in  Fig.  1. The C exhibits a distinct peak at 23 K,
presumably arising from  magnetic ordering. Above 25 K,  there  is  an
upward  curvature  of C  and  after subtraction of the lattice
contribution employing  the  values  of Y$_2$PdSi$_3$ (Ref. 8) following
the procedure suggested by Blanco et al [13], we see a broad  peak in the
data (C$_m$), which is attributed to the Schottky anomalies arising from
crystal-field effects. The low temperature $\chi$ data is plotted in the
form  of d$\chi$/dT in Fig. 1b, which shows a minimum at 23 K and at the
same  temperature there is a drop in zero-field $\rho$ (Fig. 1c).  These
results establish  that Tb ions undergo long range magnetic ordering
at 23 K (hereafter referred to as T$_c$). The  plot  of  inverse $\chi$
versus T is linear above 30 K (Fig. 1d) and  the  effective  moment
($\mu$$_{\hbox{eff}})$ obtained from the slope of the plot turns out  to
be 9.87 $\mu$$_{\beta }$ typical  of trivalent Tb ions. The Curie-Weiss
parameter ($\theta$$_p$) is found  to  be positive (11 K); while the
positive sign indicates ferromagnetic ordering, the magnitude is lower
than the actually observed value of T$_c$ (= 23 K), as if
antiferromagnetic correlations are also present.  A careful look at the
low temperature $\chi$ data (see  inset,  Fig.  1d) suggests that there is
a tendency for $\chi$ to saturate below 10 K, possibly due to another
magnetic transition. The plots (Fig. 1e) of isothermal magnetization (M)
at 2 and 10 K are not linear, which might indicate ferromagnetic
correlations;  the  saturation even at higher fields is absent; though the
plots  even above  ordering temperature (30 and 35 K) look similar to
those at 2 and 10 K presumably due to short range ferromagnetic
correlations, the increase of M for initial application of the field is
steeper at these low temperatures. The plots at 2 and 10 K look like a
superposition of a saturated ferromagnetic component and a field dependent
(antiferromagnetic) contribution. All these results indicate that  the
magnetic behavior of Tb$_2$PdSi$_3$ is quite complex and,  in particular,
the magnetic structure may not be of a simple ferromagnetic-type.
\par
Now turning to the influence of the application of H on $\rho$, the $\rho$
values are essentially unaltered by the presence of H above 60 K and hence
we show the data only in the low temperature region. However, in  the
presence of a H (say, 50 kOe, Fig. 1c),  the value  of $\rho$ gets
gradually suppressed as T is decreased, rounding off the feature due to
the magnetic ordering. This results  in  negative magnetoresistance
[$\Delta \rho /\rho = \{\rho (H)-\rho (0)\}/\rho (0)$] in the vicinity of
T$_c$,  the magnitude of which increases gradually with decreasing T,
attaining a  large value of about -16\% (for H= 50  kOe)  at 20 K;  we
will  return  to  the implications of this feature later in this article.
We  have  also  measured $\Delta
\rho /\rho$ as a function of H at selected temperatures (see Fig. 1f);
at 35 K, $\Delta \rho /\rho$ varies quadratically with initial application
of H, which indicates the existence of spin-fluctuation effects; at 5 and
10 K,  there  is  a  sudden increase in the magnitude for small values of
H, followed by flattenning  in the range 10-20 kOe with a subsequent
further increase in  magnitude  for H $>$30 kOe. This finding may be
interpreted by  the proposing that in zero field there is an
antiferromagnetic component (also inferred from M versus H plots)
undergoing metamagnetic transition with the increase of magnetic field.
The negative sign of $\Delta \rho /\rho $  is  consistent with
ferromagnetic coupling.
\par
With respect to the Dy sample, there is a clear-cut $\lambda$-anomaly
below 8.5 K (Fig. 2a), arising from magnetic ordering from the Dy
sub-lattice. The C$_m$, obtained as in the case of Tb sample, shows a
broad peak presumably  due  to crystal-field effects. The plot of
d$\chi$/dT also exhibits a dip at 8 K (Fig.  2b) at the onset of magnetic
ordering. The plot of inverse $\chi$ (Fig. 2d)  is linear above 20 K and
the value of $\mu$$_{eff}$ (10.5 $\mu$$_{\beta}$) is typical  of trivalent
Dy ions. The value of $\theta$$_p$ is very close to  zero and this
indicates  equal magnitudes  of  antiferromagnetic and ferromagnetic
correlations. There  is an upturn in $\chi$ below 7 K presumably due to
another magnetic transition as in the case of Tb sample.  The isothermal M
(Fig.  2e) tends to saturate at 2 K, which implies dominance  of
ferromagnetism at high fields; however, the value of M at 55 kOe is far
less than that expected for  free Dy  ion,  which  suggests  that  the
magnetic ordering evolves from a crystalline electric-field-split level.
With respect to $\rho$ behavior, unlike in Tb sample, $\rho$ shows an
upturn  (Fig. 2c) as the T is lowered across T$_c$;  in  the presence of
H, however, the upturn vanishes; these findings establish  [14] the
formation of magnetic superzone boundary gaps,  which  suggests  complex
nature of the magnetic structure resulting in spin density  wave
formation.  As in the case of Tb sample, $\rho$ gets depressed
continuously with decreasing T from few decades of T above T$_c$. It is
interesting to note that the magnitude of $\Delta \rho /\rho $ is quite
large at higher fields around T$_c$ (see Fig. 2f), though far above T$_c$
(say, at 35 K), it is close to zero.
\par
We have also performed ac $\chi$ measurements at two different
frequencies, 106 and 1090 Hz, on both these samples, in order to look for
possible spin-glass characteristics in these compounds. This becomes
relevant considering that both antiferro- and ferro- magnetic correlations
coexist in these alloys and that R ions form a triangular lattice [5]. The
results are shown in Fig. 3. For these  alloys, there is vitually no
difference between the data recorded at two frequencies. For Tb sample, if
at all there is a very small shift of the peak temperature with increasing
frequency, it is towards low temperatures, in contrast to the expectation
for spin-glass systems. Thus these results do not favor spin-glass
freezing in the temperature range of investigation in these alloys. This
is further confirmed by the occurence of the features due to magnetic
ordering at exactly the same temperature in all the bulk measurements
reported here (see the figures) for a given alloy which is not the case
for spin-glass systems. We also note another interesting finding in the
$\chi_{ac}$ data.  There is an additional peak at about 15 K for the Tb
case and at nearly the same temperature d$\chi$/dT exhibits a shoulder
(Fig. 1b); this finding is consistent with the proposal of another
transition around 15 K.  With respect to the Dy sample also, a broad peak
at about 6 K below which $\chi$ tends to fall could be observed, though
the transition at 8 K  appears only as a shoulder due to proximity of
these two transitions.
\par
One of the points of main emphasis is  that  the  magnetoresistance  is
negative in the vicinity of T$_c$, the magnitude of which increases as T
is decreased from a few decades of T, in contrast to the general behavior
of the influence of H  on  the  conduction  electrons  to  result  in   a
small   positive magnetoresistance. We have reported similar findings
recently  in  some Gd compounds [15],  which  were  interpreted in the
following manner: there is a formation  of polarised electron  cloud
around Gd  ion due to strong s-f exchange interaction; the mobility of
these electron clouds decreases  with decreasing temperature possibly due
to randomness of the magnetic exchange interaction, resulting  from
short-range magnetic order (and not in the sense of spin-glass freezing,
which is not found in our alloys anyway); this immobile tendency
contributes to electrical resistance as the long range magnetic ordering
temperature is approached, in a way similar to the effect of
crystallographic disorder resulting in weak electron localisation [16].
The application of  external H aligns the randomly oriented spins
(originated from short range magnetic order), which apparently facilitates
free movement of carriers. This process  of increase in conductance
resulting  in negative magnetoresistance  is qualitatively similar to the
mechanism generally proposed in the literature for giant magnetoresistance
systems, viz., the ones based on La manganite perovskites, though
double-exchange mechanism is not the one that mediates magnetic ordering
in the  rare-earth alloys.  We call this  as "magnetism-mediated  electron
localisation" (some  kind  of magnetic polaron,  which  has  been  invoked
in {it semiconducting} rare-earth compounds, viz., EuSe [10], Gd  pnictides
[11]) and EuB$_6$ [17] as well as in GaAs-based semiconductors
[18]).  To  our knowledge,  the existence of weak electron localisation
effects due to exchange interaction prior to long range magnetic order has
not  been recognised in  {\it metallic systems}.
\par
To conclude, in the compounds, Tb$_2$PdSi$_3$ and Dy$_2$PdSi$_3$, long
range magnetic order sets in at 23 and 8 K  respectively,  presumably
exhibiting  complex magnetism. The $\chi$ data show features attributable
to two magnetic transitions (the second one at about 15 and 7 K
respectively) and it might  arise from  two crystallographically
inequivalent sites [4]. It is of interest to perform neutron diffraction
measurements to find out exact nature of the magnetic structures at these
transition points.  The weak electron-localisation effects attributable to
a disorder in the exchange interaction seem to contribute to  electrical
resistivity prior to long range magnetic order, though the upturn in
$\rho$ is  not  apparent in the raw $\rho$ data, resulting in a
magnetoresistance behavior qualitatively similar to that noted for La
manganite based perovskites.  Thus, the magnitude of  magnetoresistance is
also quite large not only at lower temperatures, but also in the vicinity
of magnetic ordering temperature (due to above factor) in our compounds.
This work brings out the need to consider such magnetic precursor effects
(electron locaisation mediated by disordered exchange interaction) while
interpreting the low temperature upturns in resistivity and heat-capacity
in general even in {\it metallic} systems. This issue becomes relevant
considering controversies in understanding the origin of the apparent
non-Fermi liquid behavior in some f-electron systems [18].

 \begin{figure} 
 \caption{
  (a)  The  heat-capacity,  (b)  the  temperature   derivative   of   magnetic
susceptibility, (c) the electrical resistivity in  zero  field  and  in  the
presence  of  a  magnetic  field  and  magnetoresistance  and  (d)   inverse
susceptibility as a function of temperature  for Tb$_2$PdSi$_3$.  The  isothermal
magnetization and magnetoresistance as a  function  of  magnetic  field
are shown in (e) and (f) respectively.  In  figure  (a),  the  lattice
and  the magnetic contributions to $C$ (see tect) in Tb$_2$PdSi$_3$ are
also shown. Wherever  the  lines are drawn through the data points, these
serve as guides to the eyes, except in (d), in which the line represents
Curie-Weiss region.  In  the  inset  of Fig. (d), the low temperature
$\chi $ data is plotted to  highlight  the  features below 10 K.  }
\end{figure}

 \begin{figure} 
 \caption{(a)  The  heat-capacity,  (b)  the  temperature   derivative
of   magnetic susceptibility, (c) the electrical resistivity in  zero
field  and  in  the presence  of  a  magnetic  field  and
magnetoresistance  and  (d)   inverse susceptibility as a function of
temperature  for Dy$_2$PdSi$_3$.  The  isothermal magnetization and
magnetoresistance as a  function  of  magnetic  field  are shown in (e)
and (f) respectively.  In  figure  (a),  the  lattice  and  the magnetic
contributions to C (see text) in Dy$_2$PdSi$_3$ are also shown.  Wherever
the lines are drawn through the data points, these serve  as guides to the
eyes, except in (d), in which the line represents  Curie-Weiss region. In
the inset of Fig. (d), the low temperature $\chi $ data is  plotted  to
highlight the features below 10 K.} \end{figure}
\begin{figure}
\caption{Ac susceptibility for Tb$_2$PdSi$_3$ and Dy$_2$PdSi$_3$ at two
different frequencies in an ac field of 0.8 Oe. The approximate
temperatures of two magnetic transition points for the Dy sample are
marked by vertical arrows.}
\end{figure}
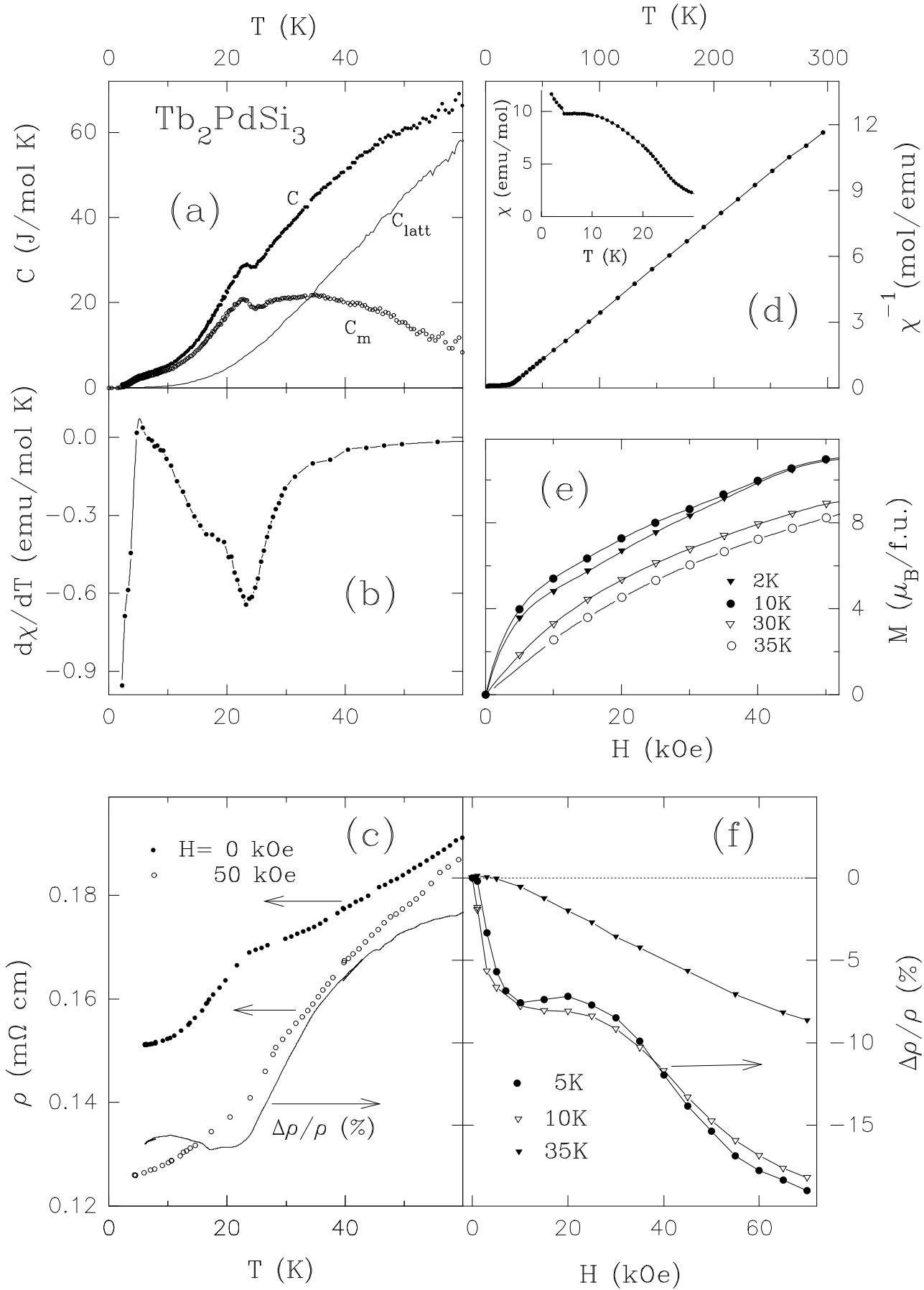
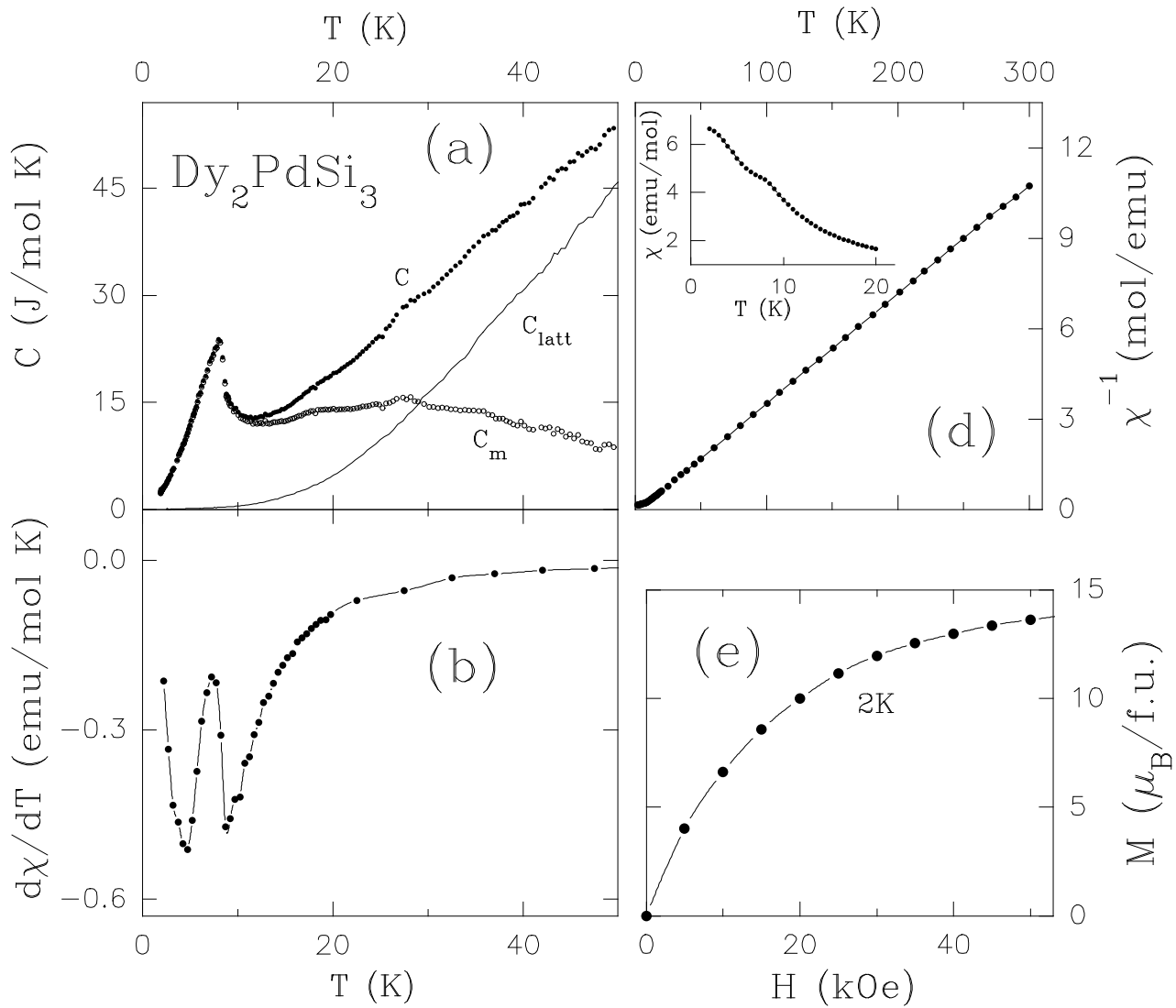
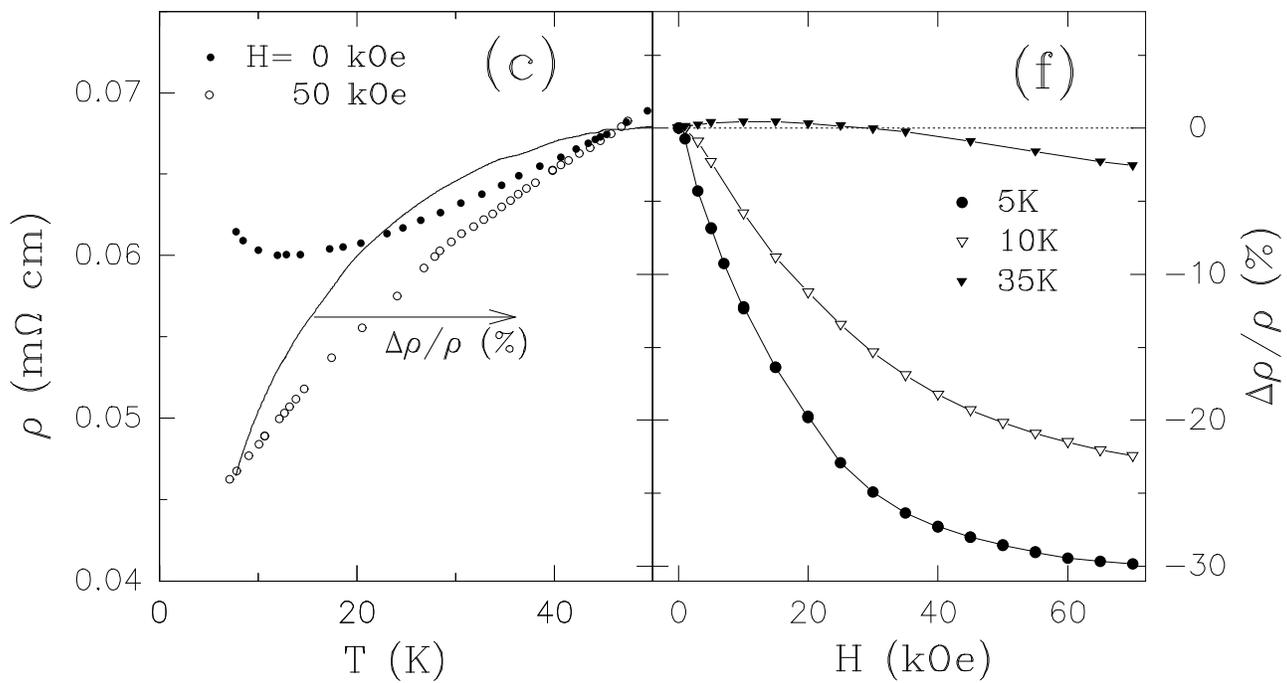
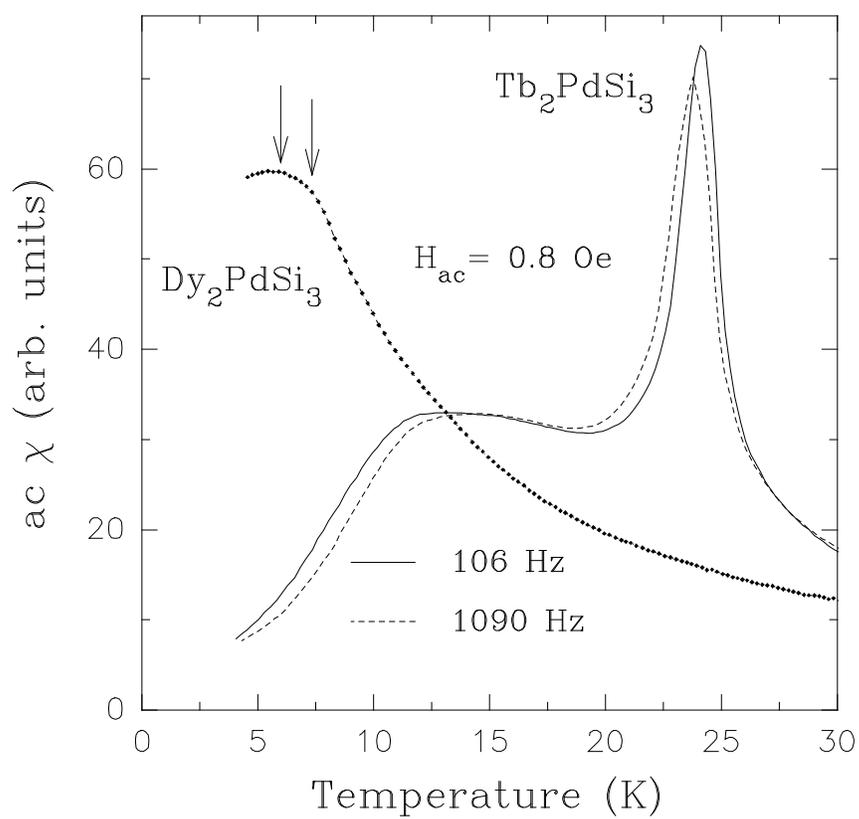
\end{document}